\documentclass[twocolumn,aps,showpacs,prb]{revtex4-1}
\usepackage{amsmath,amssymb,graphics,epsfig,epstopdf,color,multirow,array,verbatim,ulem,braket,tabularx}
\usepackage[colorlinks,linkcolor=blue,citecolor=blue,urlcolor=blue]{hyperref}
\usepackage{threeparttable}

\begin{document}

\title{LaO as a candidate substrate for realizing superconductivity in FeSe epitaxial film}

\author{Xiao-Le Qiu}
\author{Ben-Chao Gong}
\author{Huan-Cheng Yang}
\author{Zhong-Yi Lu}
\email{zlu@ruc.edu.cn}
\author{Kai Liu}
\email{kliu@ruc.edu.cn}
\affiliation{Department of Physics and Beijing Key Laboratory of Opto-electronic Functional Materials $\&$ Micro-nano Devices, Renmin University of China, Beijing 100872, China}

\date{\today}

\begin{abstract}
The significantly enhanced superconducting transition temperature ($T_c$) of an FeSe monolayer on SrTiO$_3$(001) substrate has attracted extensive attention in recent years. Here, based on first-principles electronic structure calculations, we propose another candidate substrate LaO(001) for the epitaxial growth of FeSe monolayer to realize superconductivity. Our calculations show that for the optimal adsorption structure of FeSe monolayer on LaO(001), the stripe antiferromagnetic state and the dimer antiferromagnetic state are almost energetically degenerate, indicating the existence of strong magnetic fluctuation that is beneficial to the appearance of superconductivity. According to the Bader charge analysis, the calculated electron doping from the LaO substrate to the FeSe monolayer is about 0.18 electrons per Fe atom, even larger than that in case of FeSe/SrTiO$_3$(001). Since LaO was also reported to be a superconductor with $T_c$ $\sim$ 5 K, it may have a superconducting proximity effect on the epitaxial FeSe film and vice versa. These results suggest that LaO would be an interesting substrate to study the interface-related superconductivity.
\end{abstract}

\date{\today} \maketitle

\section{INTRODUCTION}

The iron-based superconductors, being as a typical class of unconventional superconductors in addition to cuprates, have attracted intensive attention since its discovery in 2008 \cite{Y-Kamihara2008, M. Rotter2008, X-C-Wang2008, Takahashi2008, Chen-X-H2008, F-C-Hsu2008}. Among numerous iron-based superconductors, binary $\beta$-FeSe has the simplest crystal structure and provides an ideal platform to study the unconventional superconducting  mechanism. The superconducting $T_c$ of bulk FeSe is about 8 K \cite{F-C-Hsu2008}, and thus much effort has been devoted to improving the $T_c$ via high pressure \cite{Mizuguchi2008, Bendele2012}, intercalation~\cite{Guo2010, Guo2012}, chemical substitution \cite{Abdel2015}, or gate voltage \cite{Lei B2016}. In 2012, Wang \textit{et al.} reported that monolayer FeSe grown on SrTiO$_3$ (STO) substrate via molecular beam epitaxy (MBE) has the significantly enhanced $T_c$ \cite{Wang2012}, ranging from 40 K to 109 K based on different sample preparation and measurement approaches \cite{Wang2012,  He S Nat Mater 2013, Tan S Nat Mater 2013, Ge2015, Sci. Bull. 2015}. Such reports have sparked tremendous interest in the FeSe/STO interfacial system.

To explore the superconducting mechanism of FeSe/STO, many experimental and theoretical studies have been performed~\cite{WangLL2016, HoffmanJE2017, D-H. Lee2015, K-Liucpb, LeeDH2018, XuXF2020}. Some have focused on the charge degree of freedom. For example, the angle-resolved photoemission spectra (APRES) measurements have revealed that
the FeSe monolayer on STO has only electron-type Fermi surfaces at the Brillouin zone corner\cite{Nat. Commun. 2012, He S Nat Mater 2013, Tan S Nat Mater 2013}, indicating the important role of electron transfer from STO to FeSe in enhanced superconductivity~\cite{He S Nat Mater 2013, N-Li2013, Zhangcpl2014, Z-Li2014, PRB 892014, PedersenPRL}. Some have discussed the spin degree of freedom. For instance, first-principles calculations have revealed that the antiferromagnetic states of FeSe/STO are energetically lower than the nonmagnetic state\cite{PRB2012, Bazhirov2013, F-W-Zheng2013, H-Y-Cao2014}, while an \textit{in situ} ARPES measurement has suggested that the superconductivity appears when the spin density waves (SDWs) in FeSe film are suppressed\cite{Tan S Nat Mater 2013}. Such magnetic signals have also been found in later magnetic exchange bias effect experiments \cite{Zhou2018} and scanning tunneling spectroscopy measurements \cite{PRL2019}. Some other studies have highlighted the role of phonons. A study using high-resolution ARPES measurement observed replica bands in FeSe/STO~\cite{J-J-Lee2014} and attributed it to the phonon effect, which was also evidenced by an isotope experiment with $^{16}$O $\leftrightarrow$ $^{18}$O substitution\cite{Song2019}. Although the electron-phonon coupling alone cannot account for the observed superconductivity in FeSe/STO\cite{B-Li2014}, it was suggested that the coupling between FeSe electrons and STO phonons can enhance the spin fluctuations and thus the superconducting $T_c$ \cite{Y-Y-Xiang2012, S-Coh2015, K-Liu2015, Z-X-Li2016, Z-X-Li2019}. In addition, there are also other works on FeSe/STO that considered the strain effect~\cite{R-Peng2014}, the vacancy effect\cite{Bang2013, Berlijn2014}, orbital fluctuations\cite{F-Yang2013}, topological phases\cite{Hao2014}, the incipient band\cite{Linscheid2016}, nematic fluctuations and spin-orbit coupling\cite{Kang2016}, correlation strength\cite{Mandal2017}, dynamic interfacial polarons\cite{PRL1222019}, etc. These studies have largely enriched our knowledge about the FeSe/STO system.

In order to investigate the interfacial superconductivity more comprehensively, it is important to find out new substrates for the epitaxial growth of FeSe film to make a comparison with FeSe/STO~\cite{Pu2016, Wang2016}. Peng \textit{et al.} realized a superconducting gap-closing temperature of 75 K in the epitaxial FeSe monolayer on BaTiO$_3$, and they excluded the direct effect of tensile strain but highlighted the critical role of FeSe/oxide interface~\cite{Peng2014}. Rebec \textit{et al.} studied the electronic structure and superconductivity of monolayer FeSe on rutile TiO$_2$ and suggested that dielectric constant and strain are likely unimportant to increase $T_c$~\cite{Rebec2017, PRL1172016}. Zhou \textit{et al.} grew monolayer FeSe on MgO(001) substrate with an onset $T_c$ of 18 K and reemphasized the important role of charge transfer~\cite{Sci. Bull. 2019}. We notice that latest experiments have reported that the rock-salt structure LaO films deposited on YAlO$_3$(001), LaSrAlO$_4$(001), and LaAlO$_3$(001) substrates become superconducting below $\sim$5 K~\cite{Kaminaga2018}.
Notably, the in-plane lattice constant of LaO film (5.198 {\AA})~\cite{Kaminaga2018} matches well with that of bulk FeSe ($\sqrt{2}a_0=5.325$ {\AA})~\cite{F-C-Hsu2008}. As LaO has extra electrons with the usual valences of La$^{3+}$ and O$^{2-}$, it is prone to transfer electrons to the epitaxial film grown on it. Hence FeSe/LaO may be a promising system to investigate the interface-related superconductivity.

In this work, we have systematically studied the electronic and magnetic properties of an FeSe monolayer on LaO(001) substrate by using first-principles electronic structure calculations. The calculated energies of the stripe antiferromagnetic (AFM) state and the dimer AFM state for FeSe monolayer at the optimal adsorption site on LaO(001) are nearly degenerate, which may induce strong spin fluctuations that are beneficial to the emergence of superconductivity. We then studied the band structure and charge transfer of FeSe/LaO, and propose that LaO would be an interesting substrate to study the superconductivity of FeSe ultrathin film.

\section{COMPUTATIONAL DETAILS}

\label{sec:Method}
To study the atomic structure, electronic structure, and magnetic properties of an FeSe monolayer on LaO(001) substrate, fully spin-polarized density functional theory (DFT) calculations were performed with the projector augmented wave (PAW) method~\cite{Blochl1994, Kresse1999} as implemented in the VASP package~\cite{vasp1,vasp2}. The generalized gradient approximation (GGA) in the scheme of Perdew-Burke-Ernzerhof (PBE)~\cite{Perdew1996} was adopted for the exchange-correlation functional. The kinetic energy cutoff of the plane-wave basis was set to 520 eV. A six-layer LaO(001) slab with the bottom four layers fixed at their bulk positions was employed as the substrate. The in-plane lattice constants were set to the experimental values of LaO thin film on YAlO$_3$(110) ($a = b = 5.198$ {\AA})~\cite{Kaminaga2018}. A vacuum layer larger than 15 {\AA} was adopted to eliminate the interaction between image slabs along the (001) direction. The DFT-D2 method~\cite{Wu-X 2001, Grimme2006} was used to account for the possible van der Waals (vdW) interaction between layered materials. The $10\times10\times1$ and $5\times10\times1$ Monkhorst-Pack $\mathnormal{k}$-point meshes were adopted for the Brillouin zone (BZ) sampling of the $\sqrt{2}\times\sqrt{2}$ and $2\sqrt{2}\times\sqrt{2}$ two-dimensional (2D) supercells, respectively. The Gaussian smearing method with a width of 0.05 eV was used for the Fermi surface broadening. The internal atomic positions were fully optimized until the forces on unfixed atoms were all smaller than 0.01 eV/{\AA}. To study the magnetic properties of FeSe layer, the nonmagnetic state, ferromagnetic state, and four typical antiferromagnetic states (N\'eel, stripe, dimer and AFM4) were considered. For convenient comparison, the band structures of magnetic supercells were unfolded to those of unit cells by using the band unfolding method~\cite{Popescu2012} as in the PyVaspwfc package~\cite{unfold}.

\section{RESULTS AND ANALYSIS}
\label{sec:Results}

\begin{figure}[!t]
\centering
\includegraphics[width=1.0\columnwidth]{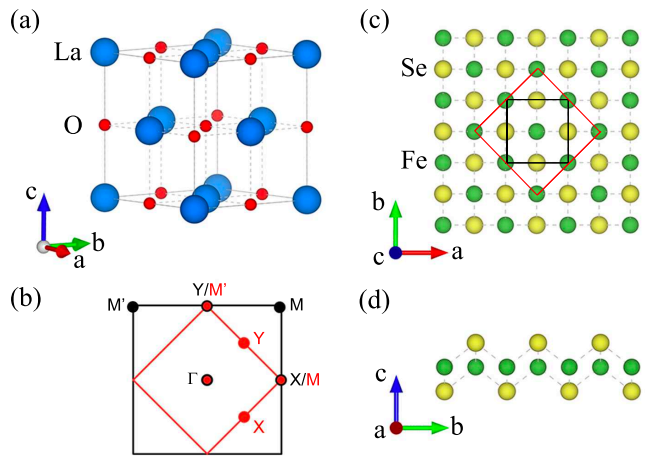}
\caption{(a) Tetragonally distorted rock-salt structure of LaO. (b) Surface Brillouin zones of the unit cell (black) and the $\sqrt{2}\times\sqrt{2}$ supercell (red) of LaO with the corresponding high-symmetry $\mathnormal{k}$ points. (c) Top and (d) side views of monolayer FeSe. The black and red squares in panel (c) outline the unit cell and the supercell, respectively. }
\label{fig:1}
\end{figure}

A Previous experiment demonstrated that the LaO(001) film deposited on the YAlO$_3$(110) surface has a tetragonally distorted rock-salt structure with the lattice constants of $a = b = 5.198$ {\AA}  and $c = 5.295$ {\AA} [Fig. \ref{fig:1}(a)]~\cite{Kaminaga2018}. These experimental values were adopted for the bulk part of LaO(001) substrate in our calculations. In fact, the in-plane lattice constant (5.198 {\AA}) of the LaO(001) film is only 2.4\% smaller than that of the $\sqrt{2}\times\sqrt{2}$ supercell of bulk FeSe~\cite{F-C-Hsu2008}, which indicates the possibility of ideal epitaxial growth of FeSe monolayer on LaO(001) substrate. Figures \ref{fig:1}(c) and \ref{fig:1}(d) show the top and side views of monolayer FeSe, where Fe atoms form a square-like plane sandwiched by two Se layers. The black and red squares in Fig. \ref{fig:1}(c) represent the unit cell and the $\sqrt{2}\times\sqrt{2}$ supercell of monolayer an FeSe, respectively. Their corresponding Brillouin zones along with the high-symmetry $k$ points are schematically shown in Fig. \ref{fig:1}(b).

\begin{figure}[!t]
\centering
\includegraphics[width=1.0\columnwidth]{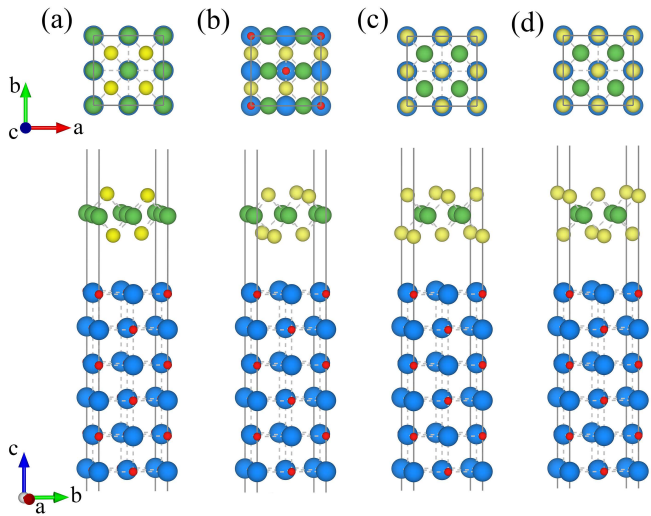}
\caption{Four typical atomic structures of an monolayer FeSe on LaO(001) surface. In each panel, the upper (bottom) part displays the top (side) view. The blue, red, green and yellow balls represent La, O, Fe and Se atoms, respectively.}
\label{fig:2}
\end{figure}

To find out the most stable epitaxial structure, we studied four possible adsorption sites for FeSe monolayer on LaO(001) surface. Figure\ref{fig:2}(a) shows the structure with Se atoms locating at the hollow sites and Fe atoms directly stacking above La or O atoms. In Fig.\ref{fig:2}(b), the FeSe monolayer shifts a quarter of cell along the $a$-axis with respect to Fig.\ref{fig:2}(a). As a result, all Fe and Se atoms are at the bridge sites in Fig.\ref{fig:2}(b). The Fig.\ref{fig:2}(c) and Fig.\ref{fig:2}(d) are constructed by shifting the FeSe monolayer in structure (b) by a quarter of cell along the negative and positive $b$-axis, respectively. We firstly studied the nonmagnetic (NM) state of FeSe monolayer and found that after structure optimization almost all above structures maintain their initial stacking patterns except for Fig.\ref{fig:2}(b), in which the optimized FeSe monolayer moves 0.6 {\AA} along the negative $b$-axis with respect to its initial site. The energetically stable patterns for FeSe monolayer in the nonmagnetic state are shown in Fig.\ref{fig:2}(a) and Fig.\ref{fig:2}(b), whose energy difference is negligible (Table \ref{tabI}).

\begin{table}[b]
\caption{Relative energies (meV/Fe) of the magnetic states (N\'eel AFM, stripe AFM, dimer AFM, and AFM4) for four epitaxial structures of FeSe/LaO(001) in Fig. \ref{fig:1} with respect to that of the nonmagnetic (NM) state in Fig.\ref{fig:2}(a).}
\begin{center}
\begin{tabular*}{\columnwidth}{@{\extracolsep{\fill}}cccccc}
\hline\hline
Structure &    NM &  N\'eel&  stripe &   dimer &   AFM4  \\
\hline
Fig.\ref{fig:2}(a)   &   0.0   &  -23.4 & -38.7 &  -38.6  &  -24.5  \\
Fig.\ref{fig:2}(b)   &   0.4   &  -12.0 & -29.9 &  -25.3  &  -20.5  \\
Fig.\ref{fig:2}(c)   &   42.0   &  28.0 & 1.2 &  8.5  &  11.0  \\
Fig.\ref{fig:2}(d)   &   195.1   &  146.3 & 138.4 &  127.9  &  153.3  \\
\hline\hline
\end{tabular*}
\end{center}
\label{tabI}
\end{table}

\begin{figure}[!t]
\centering
\includegraphics[width=0.75\columnwidth]{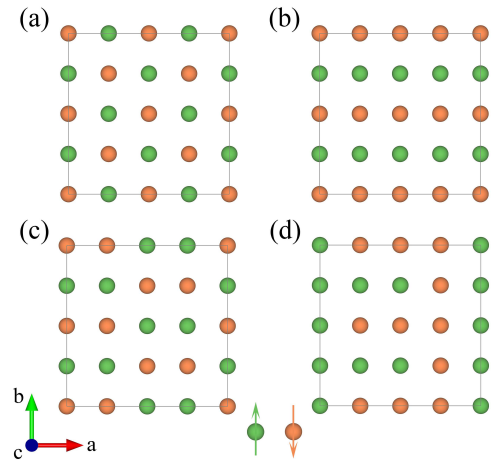}
\caption{Four typical spin configurations of Fe atoms in FeSe monolayer: (a) checkerboard AFM N\'eel state, (b) stripe (collinear) AFM state, (c) dimer AFM state and (d) AFM4 state. The green and orange balls represent the spin-up and spin-down Fe atoms, respectively.}
\label{fig:3}
\end{figure}

Many theoretical and experimental works including ours indicated that there exist magnetic interactions in FeSe/STO~\cite{K-Liucpb, PRB2012, F-W-Zheng2013, H-Y-Cao2014, Zhou2018, K-Liu2015, Wen2016}, thus we have further studied the magnetic properties of FeSe monolayer on LaO(001) substrate. In addition to the aforementioned NM state, we have further investigated the ferromagnetic (FM) state and four typical AFM states, whose spin patterns are schematically shown in Fig.\ref{fig:3}. After total energy minimization, the FM state relaxed to the NM state, indicating that the FM state is energetically unfavorable. The relative energies of these four AFM states (Fig.\ref{fig:3}) in four epitaxial patterns (Fig.\ref{fig:2}) with respect to that of the NM state in Fig.\ref{fig:2}(a) are listed in Table \ref{tabI}.
Overall, the stripe and dimer AFM states in Fig.\ref{fig:2}(a) have the lowest degenerate energy.
This suggests that there may be strong magnetic fluctuations in FeSe/LaO, which is beneficial to the emergence of superconductivity~\cite{Imai2009, Q-Q Ye2013, Suzuki2014, Lischner2015, Shishidou2018}.

\begin{figure}[!b]
\centering
\includegraphics[width=1.0\columnwidth]{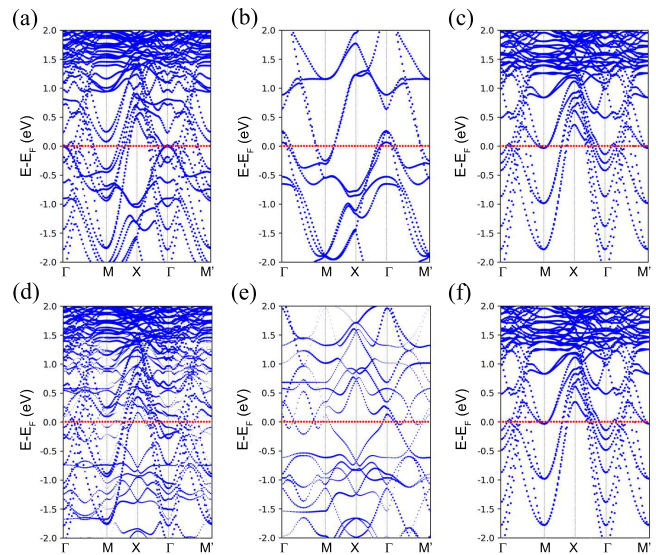}
\caption{Unfolded nonmagnetic band structures of (a) FeSe/LaO(001), (b) FeSe monolayer and (c) LaO(001) substrate. Unfolded band structures for the stripe (collinear) AFM state of (d) FeSe/LaO(001), (e) FeSe monolayer and (f) LaO(001) substrate. The $\mathnormal{k}$ paths are along the high-symmetry directions of the unit cell BZ [Fig. 1(b)]. The FeSe monolayer and LaO(001) are at their corresponding positions in FeSe/LaO(001). The red dashed lines denote the respective Fermi levels.}
\label{fig:4}
\end{figure}

\begin{figure}[!t]
\centering
\includegraphics[width=0.75\columnwidth]{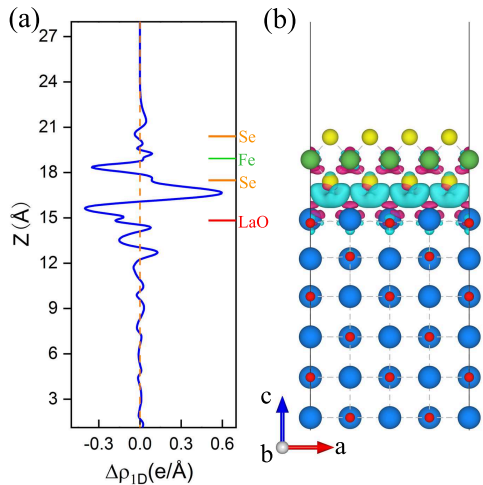}
\caption{(Color online) (a) One-dimensional and (b) three-dimensional charge difference densities for the stripe (collinear) AFM state of the monolayer FeSe on LaO(001). In panel (a), the atomic positions are marked by the color bars on the right axis. Cyan and pink isosurfaces in panel (b) are electron accumulation and depletion areas upon FeSe epitaxial growth, respectively.}
\label{fig:5}
\end{figure}

In the following, we study the electronic structures of FeSe monolayer on LaO(001). Compared with the unit cell of bulk FeSe, the supercell of FeSe/LaO is enlarged by $\sqrt{2}\times\sqrt{2}$ times (Fig. \ref{fig:2}). For the convenience of comparison, the unfolded nonmagnetic band structures of FeSe/LaO(001), FeSe monolayer, and LaO(001) substrate along the high-symmetry paths of the surface BZ of unit cell (Fig. \ref{fig:1}) are shown in Figs. \ref{fig:4}(a) to \ref{fig:4}(c), respectively. In our calculations, the structures of FeSe and LaO were fixed at their corresponding positions in FeSe/LaO(001). In general, the band structure of FeSe/LaO(001) is a superposition of those of FeSe monolayer and LaO(001) substrate with small changes in band dispersion and a shift of Fermi level [Figs. \ref{fig:4}(a)-\ref{fig:4}(c)]. For the FeSe monolayer, there are holelike bands around the $\Gamma$ point and electronlike bands around the M point [Fig. \ref{fig:4}(b)], consistent with previous study~\cite{Bazhirov2013}. After epitaxial growth on LaO(001), the holelike bands of the FeSe monolayer sink below the Fermi level [Fig. \ref{fig:4}(a)], which indicates the doping of electrons from LaO to FeSe. Furthermore, we also calculated the unfolded band structures of the stripe AFM state, as shown in Figs. \ref{fig:4}(d) to \ref{fig:4}(f). The band structure of FeSe/LaO(001) in the stripe AFM state is more complicated than the nonmagnetic one. Nevertheless, the band dispersion of FeSe monolayer with fixed atomic positions as in FeSe/LaO(001) shows an overall agreement with that of undistorted FeSe monolayer in a previous work~\cite{F-Zheng2015}, although here some bands shift to cross the Fermi level along the $\Gamma$-$M^{'}$ path. By carefully examining Figs. \ref{fig:4}(d) and \ref{fig:4}(e), we find that the FeSe bands move down after epitaxial growth on LaO(001), which is also attributed to the electron acquirement in the FeSe monolayer.

To figure out the amount of charge transfer between FeSe monolayer and LaO(001) substrate, we further calculated the one-dimensional and three-dimensional charge difference densities of the stripe AFM state as plotted in Fig. \ref{fig:5}. Clearly, the electrons accumulate in the interfacial region and redistribute around Fe atoms, which are similar to the case of FeSe/STO~\cite{K-Liu2015}. The calculated work function of clean LaO(001) surface is only 2.5 eV, comparable with that of the electride Ca$_2$N (2.6 eV)~\cite{K-Lee2013}. Such a low work function is in favor of releasing electrons, making LaO substrate as an electron donor in FeSe/LaO. Based on Bader charge analysis, there are about 0.183 (0.176) electrons per Fe atom transferred from LaO substrate to FeSe monolayer in the stripe (dimer) AFM state. In comparison, the ARPES experiment indicated that there are about 0.12 electrons per Fe atom doped to FeSe monolayer in FeSe/STO~\cite{He S Nat Mater 2013}.

\begin{figure}[!t]
\centering
\includegraphics[width=1.0\columnwidth]{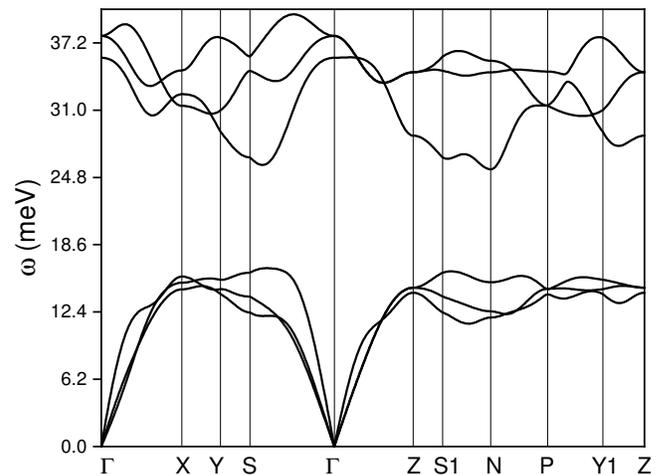}
\caption{Phonon dispersion of bulk LaO along the high-symmetry paths of the BZ of primitive cell calculated with the experimental lattice constants $a= b=5.198$ {\AA}  and $c = 5.295$ {\AA}~\cite{Kaminaga2018}. }
\label{fig:6}
\end{figure}

\section{Discussion and summary}
\label{sec:discussion}

The rock-salt structure LaO has many advantages as the substrate for the epitaxial growth of FeSe ultrathin film. First, the same in-plane tetragonal symmetry and the small lattice mismatch between the LaO(001) surface and the FeSe monolayer allow for an ideal epitaxial growth without structural reconstruction. Second, the low work function of LaO surface indicates that it is prone to induce electron doping to the adsorbed FeSe monolayer. According to the previous studies on FeSe/STO~\cite{Nat. Commun. 2012, He S Nat Mater 2013, Tan S Nat Mater 2013, N-Li2013, Zhangcpl2014, Z-Li2014, PRB 892014, PedersenPRL}, the electron doping is helpful for the enhancement of the superconducting $T_c$. Third, the oxygen optical phonon at the FeSe/LaO interface is different from that of FeSe/STO in both the frequencies (Fig. 6 in Appendix) and the induced electric dipole fields~\cite{S.-Y. Zhangprb}. This may result in different electron-phonon interactions between the FeSe electrons and the substrate phonons, thus offering an opportunity to examine the role of interfacial phonon. Last but not least, the rock-salt structure LaO was also reported to be a superconductor with $T_c$ $\sim$ 5 K~\cite{Kaminaga2018}. Hence the FeSe/LaO interfacial system might be an interesting platform for studying the superconducting proximity effect between LaO and FeSe.

The charge transfer from the LaO substrate to the FeSe epitaxial film has an important influence on the band structure of the latter. Compared with that of pristine FeSe [Fig. 4(b)], the holelike Fermi pockets of FeSe eptaxial film at the BZ center disappear and the electronlike Fermi pockets around the M point enlarge [Fig. 4(a)]. Similar changes in electronic structures have been reported for several FeSe-based superconductors with $T_c$$ >$ 30K~\cite{Nat. Mater., Front. Phys., D.-X. Mouprl,  He S Nat Mater 2013, B. Leiprb, L. ZhaoNat. Commun.}. For example,  $A_x$Fe$_{2-y}$Se$_2$ ($A_x$=K, Cs, Rb, Tl, etc.) with a $T_c$ of $\sim$30K has no holelike pocket at the $\Gamma$ point but large electronlike pockets near the M point, which challenges the Fermi surface nesting or interband scattering viewpoint in iron-based superconductors~\cite{Nat. Mater., Front. Phys., D.-X. Mouprl}. As to FeSe/STO, the superconducting phase characterized by the electronlike pockets around the BZ corner emerges along with the annealing process that induces electron doping~\cite{He S Nat Mater 2013}. More recently, (Li,Fe)OHFeSe superconductor with a $T_c$ of 43 K~\cite{B. Leiprb} was studied via the high-resolution ARPES measurements~\cite{L. ZhaoNat. Commun.}, which again only show electronlike pockets around the M point. It was proposed that the suppression of the holelike bands around the $\Gamma$ point in FeSe/STO and (Li,Fe)OHFeSe is conducive to the realization of high-temperature superconductivity~\cite{L. ZhaoNat. Commun.}. In consideration of the similar band structures with the above FeSe-based superconductors, we suggest that FeSe/LaO may also achieve high-$T_c$ unconventional superconductivity
.

Spin fluctuations are believed to play an important role in the unconventional superconductivity, so it is also crucial to study the magnetic interactions in FeSe/LaO. The magnetic coupling between Fe spins in the FeSe monolayer on LaO(001) can be understood with an effective Heisenberg model~\cite{F-J-Ma2008, Fengjie Ma2009, Liu2016},
\begin{equation}
H=J_1\sum_{<ij>}\vec{S_i}\cdot\vec{S_j}+J_2\sum_{{\ll}ij{\gg}}\vec{S_i}\cdot\vec{S_j}+J_3\sum_{{\lll}ij{\ggg}}\vec{S_i}\cdot\vec{S_j},
\end{equation}
where $J_1$, $J_2$ and $J_3$ denote the respective couplings between the nearest-, next-nearest-, and next-next-nearest- neighboring Fe spins, and $S$ is the local magnetic moment on Fe atom. According to above model, the stripe AFM state will have lower energy than the AFM N\'eel state when $J_1< 2J_2$, while the stripe and dimer AFM states will be energetically degenerate once the couplings fulfill $J_1\approx 2J_2 - 2J_3$. From our calculated energy differences among the magnetic states of the FeSe monolayer on LaO(001) (Table \ref{tabI}), we obtain $J_1 = 14.14$ meV/S$^2$, $J_2 = 9.00$ meV/S$^2$ and $J_3 = 1.93$ meV/S$^2$, which naturally explains the energy sequence and degeneracy of those magnetic states. The spin fluctuations due to the magnetic frustration between the low-energy AFM states in FeSe/LaO is conducive to the emergence of unconventional superconductivity, which waits for future experimental verification.

In summary, the electronic and magnetic properties of FeSe monolayer on  LaO(001) surface have been systematically studied by using first-principles electronic structure calculations. We find that in the most stable epitaxial structure Fe atoms in an FeSe monolayer prefer to reside on top of La or O atoms. More importantly, the stripe (collinear) AFM state and the dimer AFM state are almost energetically degenerate, indicating the existence of strong magnetic fluctuations in FeSe monolayer on LaO(001) that are likely to induce unconventional superconductivity. Because of the low work function of LaO, the amount of electrons transferred from LaO to FeSe is even larger than that from STO. In consideration of the fact that the rock-salt structure LaO itself was also reported to be a superconductor ($T_c$ $\sim$ 5 K)~\cite{Kaminaga2018}, our results suggest that FeSe/LaO  might be another interesting platform to study interfacial superconductivity in addition to FeSe/STO.

\begin{acknowledgments}

This work was supported by the National Key R\&D Program of China (Grants No. 2017YFA0302903 and No. 2019YFA0308603), the National Natural Science Foundation of China (Grants No. 11774424 and No. 11774422), the Beijing Natural Science Foundation (Grant No. Z200005), and the CAS Interdisciplinary Innovation Team. X.-L.Q. was supported by the Fundamental Research Funds for the Central Universities, and the Research Funds of Renmin University of China (Grant No. 20XNH064). Computational resources were provided by the Physical Laboratory of High Performance Computing at Renmin University of China.

\end{acknowledgments}

\begin{appendix}

\section{Phonon dispersion of LaO}

Figure \ref{fig:6} shows the phonon dispersion of bulk LaO which was obtained with density functional perturbation theory~\cite{S. Baroni} calculations as implemented in the QUANTUM ESPRESSO package~\cite{QE}.The low-frequency branch is mainly contributed by the La vibrations, while in the high-frequency branch the O vibrations play a major role.

\end{appendix}

\end{document}